\documentclass[9pt,twoside,british,a4paper,twocolumn]{article}
\usepackage[showframe]{geometry}% http://ctan.org/pkg/geometry
\usepackage{lipsum}% http://ctan.org/pkg/lipsum
\usepackage{amsmath}
\usepackage{amsfonts}
\usepackage{amssymb,upref}
\usepackage{tgheros} %% Main sans-serif font
\usepackage{tgtermes} %% Main serif font
\usepackage{anyfontsize}
\usepackage{xcolor}
\usepackage{dblfloatfix}    % To enable figures at the bottom of page
\usepackage{graphicx}
\usepackage{multicol}
\usepackage{enumitem}
\usepackage{subfig}
\usepackage{siunitx}
\usepackage{balance}
\usepackage{cite}
\usepackage{color}

\newcommand{\blue}{\color[rgb]{0,0,0.6}}
\definecolor{npblue}{rgb}{0,0,0.6}

\usepackage[T1]{fontenc}

\usepackage{newtxmath}
\usepackage{geometry}
\usepackage{fancyhdr}

\fancypagestyle{firstpagestyle}
{%

\fancyhead[RO]{\footnotesize\sffamily \LARGE\colorbox{npblue}{\color[rgb]{1,1,1} {\huge NATURE PHYSICS} \hspace*{8.8cm} LETTERS}}
\fancyfoot[L]{\scriptsize \sffamily 
\textsuperscript{1}Mathematical Oncology Laboratory, Universidad de Castilla-La Mancha, Spain; 
\textsuperscript{2}Brain Metastasis Group, Spanish National Cancer Research Centre (CNIO), Madrid, Spain; 
\textsuperscript{3}Neuro-oncology Unit, Health Institute Carlos III-UFIEC, Madrid, Spain; 
\textsuperscript{4}Department of Physics, University of Tehran, Iran;  
\textsuperscript{5}Department of Mathematics, Universidad de C\'adiz, Spain; 
\textsuperscript{6}Department of Radiation Oncology, Sanchinarro University Hospital, HM Hospitales, Spain;  
\textsuperscript{7}Department of Neuroradiology, Sanchinarro University Hospital, HM Hospitales, Spain;  
\textsuperscript{8} Inserm U1232, Centre de Recherche en Canc\' erologie et Immunologie Nantes-Angers, Nantes, F-44007, France;
\textsuperscript{9}Neurosurgery Clinic, Bern Insespital, Switzerland; 
\textsuperscript{10}Radiology Unit, MD Anderson Cancer Center, Madrid, Spain;
 \textsuperscript{11}Thoracic Surgery Unit, Hospital General Universitario de Albacete, Spain;  
 \textsuperscript{12}Nuclear Medicine Unit, Hospital General Universitario de Ciudad Real, Spain; 
 \textsuperscript{13}Fundaci\'on Instituto Valenciano de Oncolog\'{\i}a, Spain.
$^{*}$email: {\blue victor.perezgarcia@uclm.es}}
\fancyfoot[R]{}}

\usepackage{titlesec}
\titleformat{\section}[block]{\bfseries\upshape\sffamily\boldmath}{}{0.em}{}
\titlespacing*{\section}{0pt}{0.8em plus 0ex minus 0ex}{0em plus 0.ex}

\usepackage{titling}
\pretitle{\noindent\begin{minipage}{\textwidth} \raggedright\Huge\bfseries\boldmath}
\posttitle{\end{minipage} \vskip 0.6em}
\preauthor{\noindent \begin{minipage}{\textwidth} \large}
\postauthor{
  \end{minipage} 
  \vskip 0.4em
  {
   \sffamily
   \color{black!80}
   \noindent \small
   \address
   \par %\vskip -1.em
   \authoremail
   }
   \par \vskip 0.4em
  }
\predate{ \noindent \hspace{-0.4em} \sffamily\small}
\postdate{\vskip 0.0em}

\usepackage{caption}
\DeclareCaptionFont{cfs}{\fontsize{8}{10.25}\selectfont}
\DeclareCaptionLabelSeparator{vline}{\;${\boldsymbol |}$\;}
\newcommand{\figurecaption}[2]{\caption[#1]{\textbf{#1.} #2}}

\usepackage{tcolorbox}
\definecolor{abstractboxcolor}{cmyk}{0.1,0,0,0}
\newtcolorbox{abstractbox}{
  arc=0pt,
  boxrule=0pt,
  colback=abstractboxcolor,
  boxsep=0.5em,
  left=0pt, right=0pt, bottom=0pt, top=0pt,
  width=\columnwidth
}
\makeatletter
 \def\@textbottom{\vskip \z@ \@plus 1pt}
 \let\@texttop\relax
\makeatother

\makeatletter 
\renewcommand\@biblabel[1]{#1.} 
\makeatother

\usepackage[%
  bookmarks=true,
  colorlinks,
  linkcolor=blue,
  urlcolor=blue,
  citecolor=blue,
  plainpages=false,
  pdfpagelabels,
  final,
  breaklinks=true
]{hyperref}
\hypersetup{
pdftitle={Scaling laws in human cancers}, 
pdfauthor={VM P\'erez-Garc\'ia},
pdfkeywords={latex, article, class}
}

\let\OLDthebibliography\thebibliography
\renewcommand\thebibliography[1]{
  \OLDthebibliography{#1}
  \setlength{\parskip}{0pt}
  \setlength{\itemsep}{0pt plus 0.3ex}
}

\title{Universal scaling laws rule explosive growth in human cancers}

\author{\small V\'{\i}ctor M. P\'erez-Garc\'{\i}a\textsuperscript{1$\,*$}, 
Gabriel F. Calvo\textsuperscript{1}, 
Jes\'us J. Bosque\textsuperscript{1}, 
Odelaisy Le\'on-Triana\textsuperscript{1}, 
Juan Jim\'enez\textsuperscript{1}, 
Juli\'an P\'erez-Beteta\textsuperscript{1},  
Juan Belmonte-Beitia\textsuperscript{1}, 
Manuel Valiente\textsuperscript{2}, 
Luc\'{\i}a Zhu\textsuperscript{2}, 
Pedro Garc\'{\i}a-G\'omez\textsuperscript{2}, 
Pilar S\'anchez-G\'omez\textsuperscript{3}, 
Esther Hern\'andez-San Miguel\textsuperscript{3}, 
Rafael Hortig\" uela\textsuperscript{3}, 
Youness Azimzade\textsuperscript{4}, 
David Molina-Garc\'{\i}a\textsuperscript{1},  
\'Alvaro Mart\'{\i}nez\textsuperscript{1,5},
 \'Angel Acosta Rojas\textsuperscript{6}, 
 Ana Ortiz de Mendivil\textsuperscript{7}, 
 Francois Vallette\textsuperscript{8}, 
 Philippe Schucht\textsuperscript{9}, 
 Michael Murek\textsuperscript{9},  
 Mar\'{\i}a P\'erez-Cano\textsuperscript{1},  
 David Albillo\textsuperscript{10},  
 Antonio F. Honguero Mart\'{\i}nez\textsuperscript{11}, 
 Germ\'an A. Jim\'enez Londo\~no\textsuperscript{12},
 Estanislao Arana\textsuperscript{13} \& 
 Ana M. Garc\'{\i}a Vicente\textsuperscript{12}}

\newcommand{\address}{}
\newcommand{\authoremail}{}

\date{ }

\usepackage{physics}

\usepackage{textcomp}

\begin{document}
\twocolumn[
\begin{@twocolumnfalse}

\maketitle
\thispagestyle{firstpagestyle}

\vspace{-2mm}

\end{@twocolumnfalse}
]
%\fontsize{9}{11}\selectfont
\fontfamily{cmr} \fontsize{9}{10.25} \selectfont

\noindent\textbf{Most physical and other natural systems are complex entities composed of a large number of interacting individual elements. It is a surprising fact that they often obey the so-called scaling laws relating an observable quantity with a measure of the size of the system. Here we describe the discovery of universal superlinear metabolic scaling laws in human cancers. This dependence underpins increasing tumour aggressiveness, due to evolutionary dynamics, which leads to an explosive growth as the disease progresses. We validated this dynamic using longitudinal volumetric data of different histologies from large cohorts of cancer patients. To explain our observations we put forward increasingly-complex biologically-inspired mathematical models that captured the key processes governing tumor growth. Our models predicted that the emergence of superlinear allometric scaling laws is an inherently three-dimensional phenomenon.
Moreover, the scaling laws thereby identified allowed us to define a set of metabolic metrics with prognostic value, thus providing added clinical utility to the base findings.}
\par

%\textbf{Living organisms, galaxies, economies, cities and companies and many physical systems are complex entities composed of a large number of interacting individual elements. It is a surprising fact that they often obey simple laws reflecting an extraordinary simplicity when viewed as a function of their size. These so-called scaling laws are of the form  $Z = \alpha V^\beta$, where $Z$ is an observable quantity, $V$ is a measure of the size of the system -in living systems typically their volume or mass- $\alpha$ is a rate constant and $\beta$ represents the scaling exponent \cite{WestBook}. Here we describe the discovery of metabolic scaling laws in  human cancers that display superlinear exponents. This dependence underpins an increasing tumour aggressiveness due to evolutionary dynamics leading to an explosive growth as the disease progresses. We validated this dynamics using longitudinal volumetric data of distinct histologies from large cohorts of cancer patients. Our observations were explained by a number of biologically-inspired mathematical models. The identified scaling laws allowed us to define a set of metabolic metrics for each tumour having a prognostic value, thus providing a clinical added value to the fundamental findings.}
%\par
%\bigskip
%{\bf One sentence summary:} Metabolic and morphological imaging data evidences explosive growth in human cancers due to evolutionary dynamics.

%%
\begin{figure*}[!h]
\centering
\includegraphics[width=\textwidth]{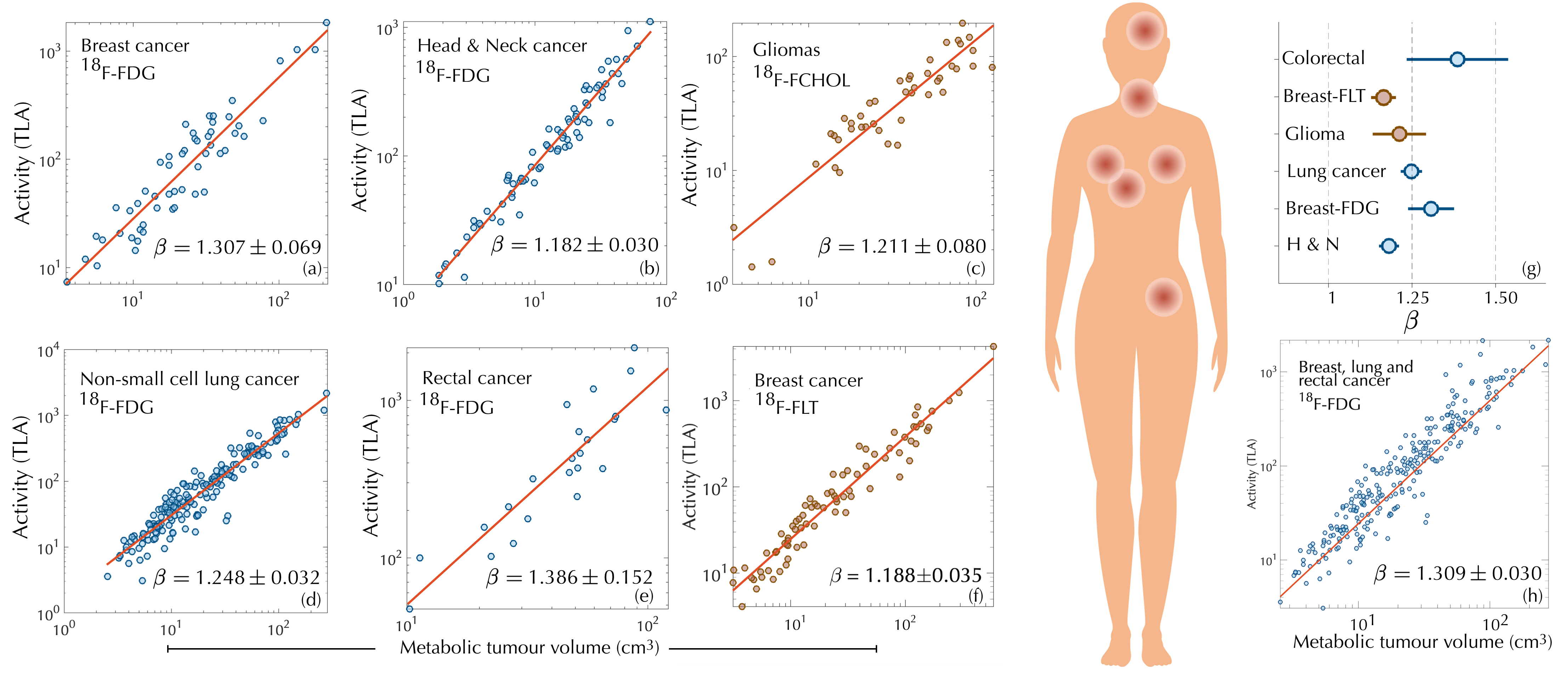}  
\centering
\figurecaption{A superlinear scaling law governs glucose uptake and proliferation in human cancers}{Log-log plots of $^{18}$F-FDG uptake (TLA) versus metabolic tumour volume on diagnostic PET for breast cancer, head \& neck cancer, non-small-cell lung cancer and rectal cancer display superlinear ($\beta > 1$) allometric scaling laws. Diagnostic PET with proliferation radiotracers, either $^{18}$F-FLT for breast cancer or $^{18}$F-FCHOL for glioma, shows the same dependence pointing to the use of glucose mostly as a resource for biosynthesis. The fitted exponents cluster around $\beta = 5/4$. Joint records of patients imaged in the same institution with identical protocol (breast-FDG, lung and rectal cancers), show that a common scaling law governs the dynamics. Error bars in (g) correspond to the standard error in the fitted parameter $\beta$ obtained using \texttt{fitlm}.}
\label{scaling}
\end{figure*}

Biological systems display complex spatially and temporally varying structures that are mainly a consequence of their underlying metabolism. Organisms continuously incorporate energetic and material resources from the environment, transforming and allocating them into different compartments that allow for their growth, reproduction and, hence, survival, both as individuals and as species. Metabolism involves random fluctuations and hierarchical processes that determine the pace at which organisms live and evolve.  In a seminal work \cite{Kleiber}, Kleiber observed that, for a broad variety of species, metabolic rates scale to the 3/4 power of the animal's mass. This result contradicted theories assuming a direct proportionality between the animal's volume and its metabolic rate, or other scalings such as metabolic rate being proportional to the animal surface. Scaling laws are of the form  $Z = \alpha V^\beta$, where $Z$ is an observable quantity, $V$ is a measure of the size of the system -in living systems typically their volume or mass- $\alpha$ is a rate constant and $\beta$ represents the scaling exponent \cite{WestBook}. West and coworkers proposed that the exponent $\beta = 3/4$ found by Kleiber could be the result of principles of minimal energy \cite{WBB}. Many related studies have explored allometric scaling laws in other biological contexts\cite{R1,Reich,R4}.

\par

Do human cancers obey metabolic scaling laws? Some evidence obtained from {\em in vitro} experiments or from xenotransplantation of patient-derived cells into immunocompromised mice seem to support that cancers also obey the Kleiber's law or similar sublinear dynamics\cite{UniversalLaw,Herman,MetabScal}. However, no works have uncovered scalings laws from large cancer patient datasets. Here we addressed this question under the initial hypotheses that malignant tumours would scale between the metabolic requirements of coordinated tissues governed by minimal energy principles (leading to an exponent $\beta \simeq 3/4$) and that of independent uncoordinated units (exponent $\beta \simeq 1$).

\par

Tumour cells exhibit high metabolic requirements to sustain an upregulated proliferation. Nutrients such as glucose and, to a lesser extent, glutamine are mostly used to fuel biomass formation and macromolecule synthesis\cite{Palm}. Deregulated glucose uptake by tumour cells, known as the Warburg effect, constitutes the basis of positron-emission-tomography/computed-tomography (PET/CT)-based imaging by means of the radioactive tracer $^{18}$F-fluorodeoxyglucose ($^{18}$F-FDG), widely used in clinical oncology\cite{Zhu2019}. To study the relationship between tumour metabolic rates and volume we collected data of different cancer types imaged at diagnosis with $^{18}$F-FDG PET/CT. Tumours were segmented and their total lesion activity (TLA) and metabolic tumour volume (MTV) calculated. TLA and MTV were computed as the product of each voxel volume within the tumour by its measured standardized uptake value (SUV) and as the summed volume of the segmented tumour voxels, respectively. Our first goal was to determine whether a dependence of the form $\textrm{TLA}\sim\alpha \textrm{MTV}^{\beta}$ could be identified. Figure \ref{scaling} shows log-log plots of MTV versus TLA for patients with: locally advanced breast cancer (LABC), head and neck cancer (H\&NC, stages II-IV), non-small-cell lung cancer (NSCLC, stages I-III) and rectal cancer  (RC, stages III-IV) (see `Methods' for more patient data). The obtained exponents were,  
$\beta = 1.307 \pm 0.069$ (R$^\text{2}$ = 0.874, LABC), 
$\beta = 1.182 \pm 0.030$ (R$^\text{2}$ = 0.954, H\&NC),
$\beta = 1.248 \pm 0.032$ (R$^\text{2}$ = 0.900, NSCLC),
 $\beta =  1.386 \pm 0.152$ (R$^\text{2}$ = 0.798, RC), as shown in Fig. \ref{scaling}(a,b,d,e). 
Thus, superlinear scalings  clustered around the rational number $\beta = 5/4$ (Fig. \ref{scaling}(g)).  Moreover, all patients scanned in the same institution and undergoing an identical protocol, thus providing comparable data, followed a common scaling law with $\beta = 1.309 \pm 0.030$ (R$^\text{2}$ = 0.895) [Fig. \ref{scaling}(h)]. Possible artefacts on the scaling exponents due to the partial volume effect in PET images were discarded. Our findings contradicted the hypothesis of metabolic scaling being sublinear suggesting a fundamentally different dynamics.

\par

This superlinear glucose uptake could be the result of different mechanisms. The first one would be an increase of the Warburg phenotype leading to a less efficient use of glucose. Also, the presence of immune cells and inflammation within the tumour region could be a contributing factor. However, since glucose is mostly used to satisfy the proliferation demands \cite{Palm,Zhu2019}, we suspected that an increase of the proliferation rate with size was probably the main underlying cause.

To clarify this, we gathered data from glioma patients (grades II-IV) imaged at diagnosis with $^{18}$F-Fluorocholine PET ($^{18}$F-FCHOL), and from  breast cancer patients (stages II-IV) imaged at diagnosis with 3'-deoxy-3'-$^{18}$F-fluorothymidine  PET ($^{18}$F-FLT). These two radiotracers reflect choline and thymidine metabolism and are related to cell proliferation\cite{Choline,FLT}. The obtained scaling exponents were $\beta = 1.21 \pm 0.08$ for gliomas and $\beta = 1.188 \pm 0.035$ for breast cancers (Fig. \ref{scaling}(c,f)), in agreement with a superlinear activity and providing support to the hypothesis of an increased glucose uptake to satisfy the proliferation demands. 

\par

\begin{figure*}[t]
\centering
\includegraphics[width=\textwidth]{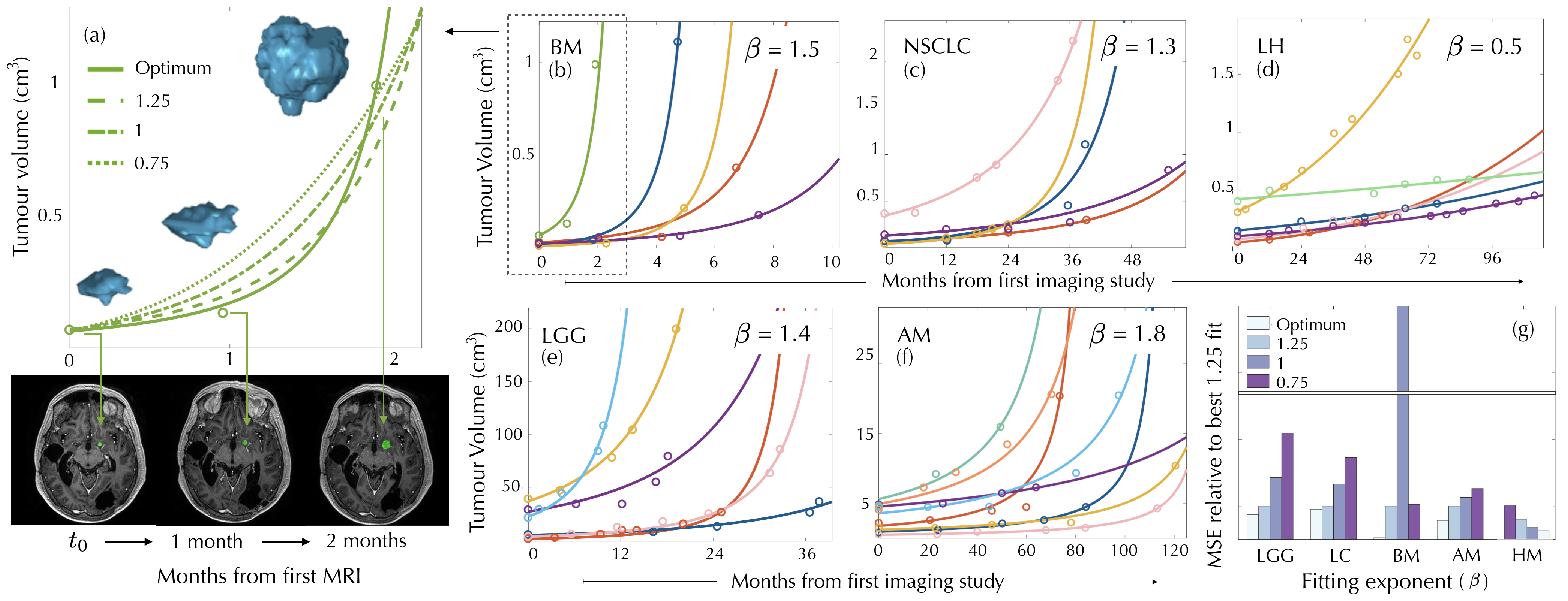}
  \centering
\figurecaption{Explosive longitudinal volumetric dynamics of untreated malignant human tumours}{Longitudinal volumetric data for cancer patients with untreated brain metastases (BM), low grade gliomas (LGG), non-small-cell lung carcinomas (NSCLC), atypical meningiomas (AM) and lung hamartomas (LH). Solid curves show the fits with the optimal exponents (values provided in each subplot) giving the smallest mean square errors. The longitudinal 3D reconstruction of a BM and representative axial slices highlighting tumour location at three time points are displayed in the left panel together with the fitting curves obtained for different exponents. Mean square errors (MSE) for the five datasets and exponents $3/4$, $1$, $5/4$ (taken as a reference) in comparison with the optimal exponent, are depicted in the lower right subplot.} 
\label{explosion}
\end{figure*}

Superlinear scaling laws have been found in varied scenarios, ranging from urban infraestructures and socioeconomic networks to primitive life forms \cite{WestBook}. In contrast with sublinear scaling, which leads to stable bounded growth, superlinear scaling produces unbounded growth. For biological organisms, whole-body metabolic rates increase with size across prokaryotes, protists and metazoans, although each group is characterized by a distinctive scaling relationship that is unique to their body size range\cite{DeLong}. In heterotrophic prokaryotes the relationship between metabolic rate and body mass has an exponent $\beta > 1$, whereas for metazoans it is $\beta < 1$. Within an evolutionary perspective, the transition from simple prokaryotes to complex eukaryotes has shown not only a higher level of multicellular organization, but also a trend towards the $3/4$ scaling exponent of Kleiber's law. Our results suggest that human cancers, as they progress, decrease the efficiency of their local vascular network~\cite{DePalma}, which would tend to increase their scaling exponents and significantly deviate from the $3/4$ Kleiber's law. 
\par

To further quantify the relationship between tumour size and metabolism, let $B \propto V^{\beta}$ denote the metabolic rate of a tumour, where $V$ is the volume occupied by viable cells. 
A simple mathematical model accounting for energy conservation describing the temporal dynamics of tumour growth is $B = a V + b \frac{dV}{dt}$, where the first and second terms correspond to cell maintenance and proliferation, respectively\cite{WestLaw}. If most of the energy is used for cell biosynthesis, we may write
\begin{equation}
\label{themodel}
\frac{dV}{dt} = \alpha V^{\beta}.
\end{equation}
When $\beta > 1$, there is a finite time $t_{\text{crit}} = t_0 + V_{0}^{1-\beta}/\left[\alpha(\beta - 1)\right]$ at which the tumour 'blows up', where $V_{0}$ is the volume at time $t_{0}$. For detais see Supplementary Information (SI) Section S1. Thus, the existence of a superlinear scaling law between proliferation and volume implies an increasingly accelerated volumetric growth and the formation of a singularity in a finite time. In real cancers such dynamics cannot be sustained to the blow-up point, since tumours are subject to physical and nutrient-supply constraints. In patients, such an accelerated growth in the final stages entails metabolic and spatial requirements incompatible with life.
\par

There has long been discussion about the best mathematical model for describing tumour growth, most of them assuming different types of bounded dynamics \cite{UniversalLaw,Gerlee,RB2013,Benzecry,Talkington}. The data supporting these models comes from patient-derived cell lines cultured {\em in vitro} or, else, from either allotransplantation of murine cells into syngeneic immunocompetent inbred mice or from xenotransplantation of patient-derived cells into immunocompromised mice. These models have a number of shortcomings when compared with their human counterparts. They display loss of genetic heterogeneity and irreversible changes in gene expression due to long-term {\em in vitro} propagation\cite{Gengenbacher} and exhibit a rapid non-autochthonous growth that results in a perturbed tissue architecture with alterations in the vascular, lymphatic and immune compartments.  
\par

To investigate whether explosive tumour growth could be observed in cancer patients, we looked for longitudinal imaging datasets of untreated tumours. Data of this type is scarce since growing tumours are typically either treated or -as in the case of palliative care patients- not followed up by imaging. Most available datasets had either incomplete information, no volumetric imaging and/or very few time points. Mandonnet and colleagues\cite{Mandonnet} studied the growth dynamics of untreated WHO grade II gliomas, Van Havenbergh\cite{VanHavenbergh} analysed petroclival meningiomas, and Heesterman et al\cite{Heesterman} head and neck paragangliomas. Growth dynamics consistent with sublinear scalings were observed for those slowly growing tumours. To further confirm this idea, we collected longitudinal volumetric growth data from a set of lung hamartomas, the most frequent benign lung tumour type, and found a best fit of Eq. (\ref{themodel}) with $\beta = 0.5\pm 0.2$ (Fig. \ref{explosion}(d)). Hence, not all human tumours manifest an explosive growth.
\par

We also collected imaging datasets of patients bearing tumours that were either malignant initially or became malignant over the course of the disease (see `Methods' for a description of the patient datasets). The first one was a set of brain metastases in which one of the lesions was either below target definition or left without therapy due to medical reasons. A second set comprised initially WHO grade II gliomas that underwent surgery and then received no other treatment for long periods. The third was a set of patients enrolled in a lung cancer screening program. After detection of lung nodules with no signs of malignancy they where followed up by low-dose CT scans. Many of these tumours accelerated their growth until a point at which further therapeutical actions were taken. Finally, we included a subset of petroclival meningiomas that showed signs of atypical behaviour (cases 5, 6, 9, 11 of Fig. 7 and cases 14, 18 of Fig. 8 in Ref.~\cite{VanHavenbergh}). For each patient we fitted the longitudinal volumetric growth data using different power-law models expressed by Eq. (\ref{themodel}). We tested the  exponents $\beta = 3/4$ (size-limited Kleiber's law), $\beta =1$ (exponential growth law) and then superlinear $\beta = 5/4$. Subsequently, we searched for the exponent that minimised the mean square error (MSE) for all patients within each tumour type. In all these examined cases, the existence of an explosive growth dynamics was confirmed [Fig. \ref{explosion}(b,c,e,f)]. A comparison of the MSEs for the different exponents and tumour types is shown in Fig. \ref{explosion}(g). We also performed a least-squares fitting of the $\alpha, \beta$ parameters for each patient and computed the mean and standard deviation for patients of each pathology. The results obtained were 1.493$\pm$0.0197 (BMs), 1.360$\pm$0.2922 (NSCLC), 1.466$\pm$0.269 (LGGs) and 1.690$\pm$0.452 (AMs) respectively. Thus, exponents obtained using the two methodologies were compatible between them and superlinear.
\par

To determine whether animal models could also provide evidence of super-exponential tumour growth dynamics, we performed experiments on two animal models chosen because of their close relationship to their human counterparts. First, we injected the human lung adenocarcinoma brain tropic model H2030-BrM\cite{Nguyen} into the heart of nude mice in order to induce the formation of brain metastasis from systemically disseminated cancer cells. The exponent best fitting the dynamics of the brain metastasis measured using bioluminiscence, assuming a dynamics ruled by Eq. (\ref{themodel}), and data from all the mice was $\beta = 1.3$. The total tumour load in the animals showed similar behaviour, with $\beta = 1.25$ (Extended Data Fig. S1). In a second set of experiments, we injected primary glioma cells closely resembling the dynamics observed in patients \cite{STM} and expressing the luciferase reporter gene into the brains of nude mice. One month after the injection, weekly monitoring of the animals was started, measuring the total flow to assess tumour growth. The optimal exponents obtained were also $\beta = 1.25$.  
\par

Thus, a sustained increase in proliferation is supported both by the allometric scaling laws and the morphological longitudinal growth data during the tumour's natural history.  We suspected that evolutionary dynamics could be the underlying process. Via genomic instability, driver gene mutations can confer, to subpopulations of clonal cells, somatic fitness advantages over other cells within the same tumour, and contribute to higher proliferation rates. Mutational events are expected to occur locally in space and time. However they require time to consolidate over the whole population\cite{Hallatschek}, thus leading to an effective continuous change in the tumour's global proliferation rate. Phenotypic variability, manifested as trait fluctuations within identical genotypes, also leads to further selection of more proliferative cells\cite{Deforet}.

\par

\begin{figure*}[h]
\vskip 6mm
\begin{minipage}[c]{0.65\textwidth}
\includegraphics[width=\textwidth]{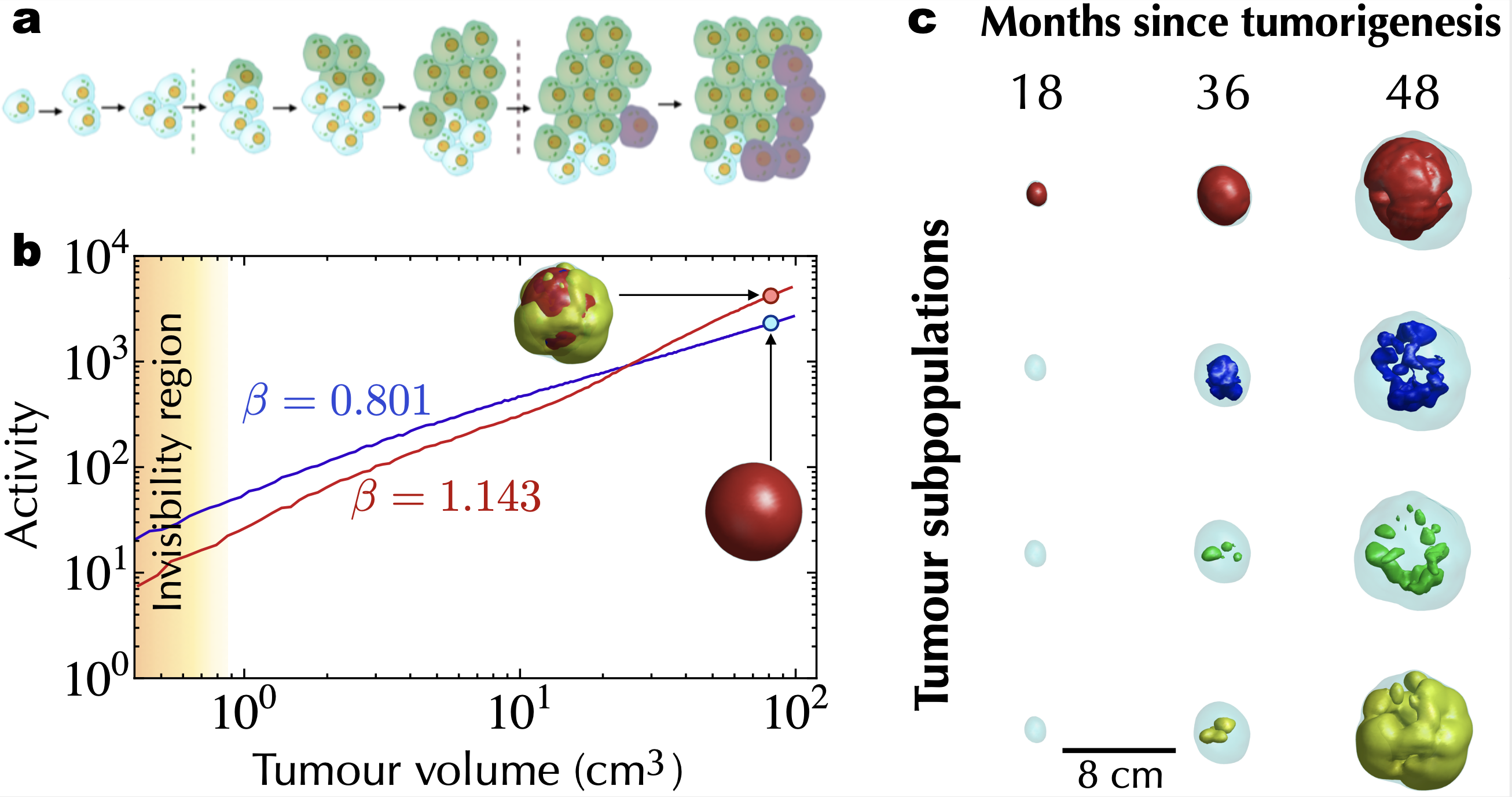}
\end{minipage} \hskip 3mm
\begin{minipage}[c]{0.32\textwidth}\vskip 2mm
\figurecaption{Stochastic mesoscale models with evolutionary dynamics lead to superlinear scaling laws {\em in silico}}{{\bf\normalsize a,} Schematic representation of the evolutionary dynamics included in the mesoscale tumour growth simulator model. Random time-local discrete events accounting for either mutations and/or phenotypic changes provide a competitive advantage to newly arising subpopulations. {\bf\normalsize b,} When a single tumour population is present, it grows continuously and displays a sublinear scaling law (blue line). In contrast, the evolutionary dynamics of a heterogeneous tumour (here consisting of four subpopulations, see SI section S3) yielded superlinear growth dynamics (red line). {\bf\normalsize c,} Isosurfaces of four interacting cell subpopulations at different points in time showing the dynamics of dominance by the most aggressive cells (higher indices correspond to more aggressive clones as described by the model parameters).} 
\label{superlineal}
\end{minipage}
\end{figure*}

The phenomenological model given by Eq. \eqref{themodel} lacks key hallmarks of real cancers.  We explored {\em in silico} increasingly sophisticated spatio-temporal models incorporating cell migration and competition among different cell subpopulations. The first mathematical model that we put forward was a nonlocal Fisher-Kolmogorov equation (NLFK), encompassing random diffusive tumour cell motion and proliferation with saturation when reaching the local carrying capacity. The NLFK reads as
\begin{equation}
\frac{\partial u}{\partial t}=D \nabla^2u+ \left( \rho_{0} + \rho_{1}N(t)\right) \left( 1 -\frac{u}{K}\right) u\, ,
\label{eq:NLFK}
\end{equation} 
where $u=u({\bf x},t)$ denotes the tumour cell density, being a function of space ${\bf x}$ and time $t$. The model parameters are: the cell diffusion constant $D>0$, the size-independent $\rho_{0}>0$ and size-dependent $\rho_{1}\geq 0$ proliferation rates and $K$ the local carrying capacity of the medium. The proliferation term in Eq. (\ref{eq:NLFK}) includes a dependence on the total number of tumour cells $N(t) = \int u({\bf x},t)\, d^{3}{\bf x}$ on the grounds that, as total tumour size increases, there will be a higher probability of accumulated mutational events leading to more aggressive clones (see SI Section 2 for a derivation of the NLFK). The proliferation activity of the tumour, in the context of this model, is given by $M(t) = dN/dt$ and yields the scaling laws. 
\par
To quantify the role of spatial dimensionality $d$ on the tumour growth scaling laws, we performed a mathematical analysis of Eq. (\ref{eq:NLFK}) [see SI Section S2]. If $\rho_{1} = 0$, one recovers the local FK equation for which the scaling exponent of $M(t)$ is $\beta = (d-1)/d < 1$, thus resulting in a sublinear  growth. When $\rho_{1} > 0$, the proliferation activity exhibits a superlinear scaling $\beta = 2 - 2/d$, leading to an explosive tumour growth only if $d=3$. The tumour radial velocity, which is a relevant metric in the clinic, can also be obtained in closed form as $v_{d}(t)=M(t)/C_{d}N^{(d-1)/d}(t)$, where $C_{1} = 2$ (1D),  $C_{2} = \left( 4\pi\right)^{1/2}$ (2D), and  $C_{3} = \left( 36\pi\right)^{1/3}$ (3D). Hence, dimensionality plays an essential role in the emergence of superlinear allometric laws within the NLFK model Eq. (\ref{eq:NLFK}).
\par

To further elucidate the contribution of different interacting cell subpopulations to the global tumour dynamics, we developed a stochastic mesoscale tumour growth simulator enabling cells to undergo replication, apoptosis, migration to neighbouring voxels and genotypic/phenotypic transitions (see SI Section S3). By mesoscale we refer to a coarse-grained approach that can reach computationally clinically relevant tumour sizes ($\sim 10^{2}$ cm$^{3}$) by working at the population level rather than on individual cells. Extensive {\em in silico} simulations showed superlinear  scaling in broad regions of the parameter space, matching both the volume range and time kinetics observed in patients (Fig. \ref{superlineal}). Superlinear behaviour was present in so far as there was a persistent overtaking of cell subpopulations by more aggressive ones. The dynamics of uniform populations, without {\em in silico} evolutionary dynamics, displayed sublinear scalings (Fig. \ref{superlineal}). Other mathematical models incorporating short-range dispersal and cell turnover have reported changes in spatial growth due to the underlying evolutionary dynamics\cite{ED1,ED2}.

\par

Scaling laws are very intriguing properties of physical and biological systems that shed light on their dynamics. They have a fundamental value but are often of limited applicability. We hypothesized that, once a scaling law of the form $Z = \alpha V^{\beta}$ is set as a reference for a specific cancer type, those with radiotracer uptake higher than the reference level, as defined by the scaling law, could be more aggressive than those with lower activity. Thus, we computed the distance with respect to a reference scaling law (DSL) for each tumour $j$ and dataset for which survival information was available, via $\text{DSL}_j = \text{TLA}_j - \alpha \text{MTV}_j^{\beta}$, and compared two sets with different DSL for the whole range of values of the prefactor $\alpha$, as described in Methods.  Figure 4 summarizes our results for a fixed exponent $\beta = 5/4$ in four patient cohorts with distinct cancer types. We found ranges of threshold values classifying patient subpopulations into DSL groups with survival differences as measured by the Harrell's c-index. 

\par

The classical metabolic variables MTV and TLA classified gliomas (MTV: c-index = 1.0, $p =$ 0.013; same for TLA) and breast cancer patients (MTV: c-index = 0.824, $p =$ 0.098; TLA: c-index = 0.87, $p =$ 0.01) but neither lung cancer nor head and neck patients. Hence,  the superlinear metabolic scaling laws provided prognostic metrics that were more robust than other classical PET-based indices.
\par

\begin{figure}[t!]
\centering
\includegraphics[width=\columnwidth]{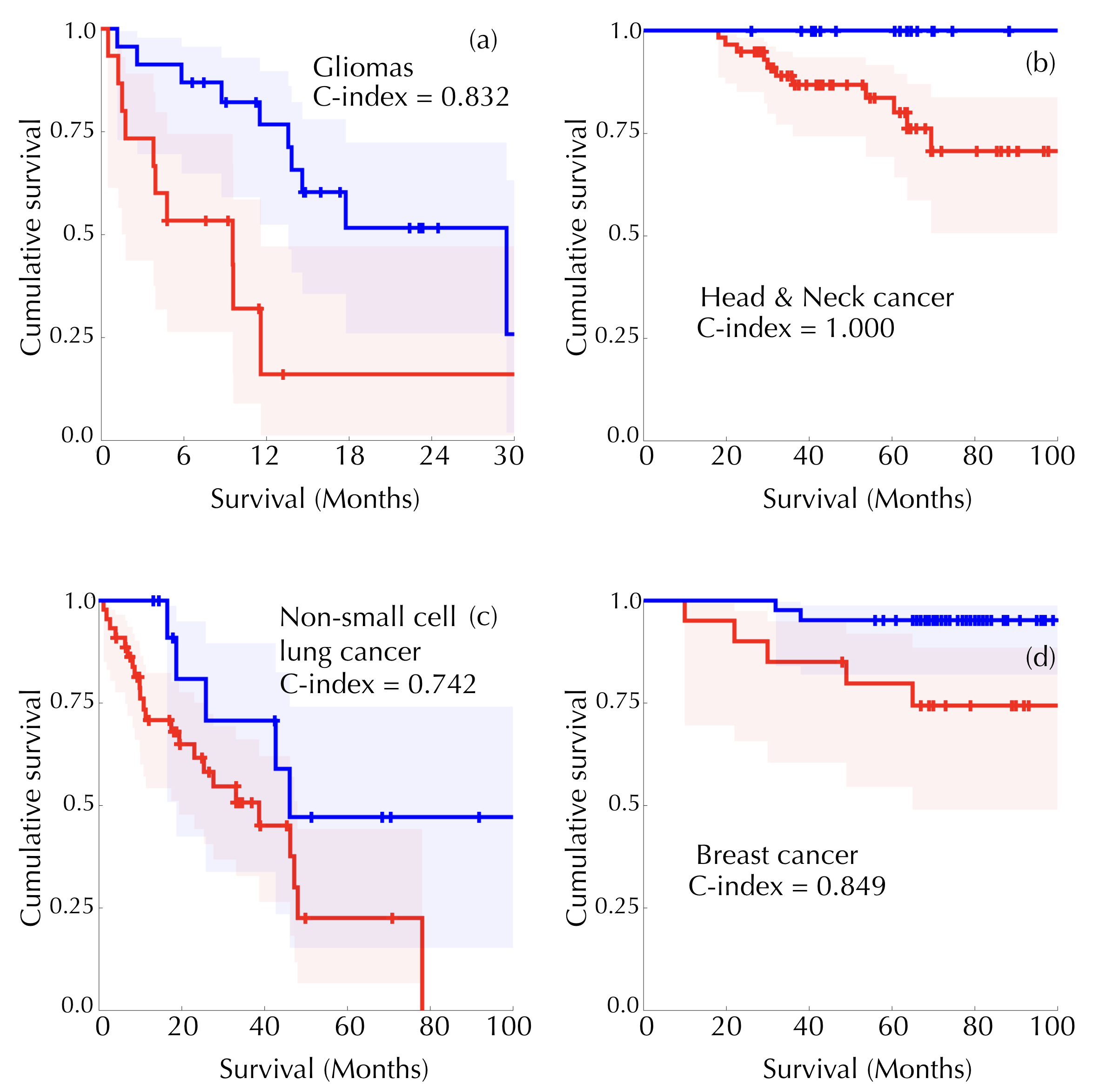}
\figurecaption{%
Scaling laws allow for cancer patient classification into prognostic groups}{
Patient tumours were classified as hyperactive (TLA > $\alpha V^{\text{5/4}}$; DSL > 0) or hypoactive (TLA < $\alpha V^{\text{5/4}}$; DSL $<$ 0) using the metabolic scaling law as a reference. Survival differences between groups were compared using Kaplan-Meier analysis and the c-index. Shown are Kaplan-Meier survival curves and the best c-index values obtained for:  
(a) Gliomas ($p=$0.001, c-index = 0.832, $\alpha =$ -0.24867). (b) Head and Neck cancer ($p=$0.05, c-index = 1.0, $\alpha =$ -0.0041776). (c) Stage III and IV resectable lung cancer patients ($p$=0.09, c-index = 0.742, $\alpha =$ -0.40334). (d) Breast cancer ($p=$0.019, c-index = 0.849, $\alpha = $-0.65034).}\label{prognosis}
\end{figure}

The observation of superlinear metabolic scaling laws and explosive behaviour of malignant tumours opens up many avenues of research. Our stochastic mesoscopic framework showed how evolutionary dynamics leads to superlinearity through the competition and consolidation of different tumour subpopulations. However, evolutionary steps could be based on mutations or phenotypic variability. When an initial driver mutation appears locally in space, even when it is more advantageous, it requires some time to consolidate. During this time window our simulations showed a continuous acceleration due to the fact that an increasing number of cells bear this new genotype. However, once this mutation is consolidated, a plateau could develop, provided no new driver mutations have appeared in the meantime. For our choice of parameters the effective dynamics resulting from our discrete simulations was in general superlinear, in agreement with our observations based on patient and animal experimental data. 

The specific mechanisms leading to an increase in the proliferation with the tumor physical size could differ between types of cancers. Some of them could be of evolutionary nature related to genotype or phenotype changes as discussed before, involve the random selection of higher fitness values\cite{Michor}, as well as to the possibility of acquiring drivers before deleterious passengers \cite{McFarland}, etc. Other potentially relevant processes 
arise in the interplay of glycolysis and tumour vascularization and oxygenation, such as the onset of the Warburg effect induced by hypoxic episodes. Others could be related to changes in the interaction between the tumour and the surrounding tissue, the action of the immune system, alterations in the tumour microenvironment such as acidosis \cite{Anderson}. Some of these effects, while possibly driven by mutational alterations, may in fact be ecological in nature. It is interesting to point out that small tumors below the spatial scale studied here, may have superlinear behaviour due to different reasons: Allee effect models in ecology have decreased growth rates at smaller tumor sizes, and these produce growth curves that are potentially indistinguishable from superlinear growth laws when fitting a few points from data. Other size-related effects for small tumors may involve the interaction with the immune system: small tumors may struggle to outgrow the immune system at first, but once they creep up to a large enough size, the immune death would become negligible.  

Our results emphasise the need to gain a better understanding of the evolutionary steps in different tumour histologies and to target these transformations to avoid growth acceleration. They also raise the question of whether working with experimental tumour models that show slower than superexponential growth could miss essential features of cancer dynamics. Finally, the role played by allometric scaling laws in human cancers under different therapies and the ultimate development of resistances has not yet been explored.
\par

In summary, we have found superlinear metabolic scaling laws in human cancers. These laws differ substantially from the Kleiber's law  governing many life forms, and point to accelerated growth due to underlying evolutionary dynamics selecting more aggressive subpopulations. Longitudinal volumetric data from malignant tumours shows explosive growth beyond classical growth-limited or exponential laws. Our mathematical models, assuming intrinsic evolutionary dynamics, put forward a mechanistic explanation for the observed phenomenology and predict that the emergence of superlinear scaling laws is an inherently three-dimensional phenomenon.
\par

\balance
\renewcommand{\refname}{}

\fontfamily{lmss} \fontsize{8.0}{10.25} \selectfont

\newcounter{mine}

\bigskip
\noindent{\bf\normalsize References}
\par

%\hfill % hacky fix for a bad-box warning caused by the balance package.

\fontfamily{lmss} \fontsize{8.5}{10.25} \selectfont

\noindent \textbf{Acknowledgements.} 

\noindent This research has been supported by the James S. Mc. Donnell Foundation 21st Century Science Initiative in Mathematical and Complex Systems Approaches for Brain Cancer (Collaborative awards 220020560 and 220020450), 
Ministerio de Econom\'ia y Competitividad/FEDER, Spain (grant number MTM2015-71200-R), Junta de Comunidades de Castilla-La Mancha (grant number SBPLY/17/180501/000154). Research in the Brain Metastasis Group is supported by MINECO grants MINECO-Retos SAF2017-89643-R (M.V.), Bristol-Myers Squibb-Melanoma Research Alliance Young Investigator Award 2017 (498103) (M.V.), Beug Foundation’s Prize for Metastasis Research 2017 (M.V.), Fundación Ramón Areces (CIVP19S8163) (M.V.), Worldwide Cancer Research (19-0177) (M.V.), H2020-FETOPEN (828972) (M.V.), Fundació La Marató de tv3 (141), Clinic and Laboratory Integration Program CRI Award 2018 (54545) (M.V.), AECC Coordinated Translational Groups 2017 (GCTRA16015SEOA) (M.V.), LAB AECC 2019 (LABAE19002VALI) (M.V.), La Caixa INPhINIT Fellowship (LCF/BQ/IN17/11620028) (P.G-G.), La Caixa-Severo Ochoa International PhD Program Fellowship (LCF/BQ/SO16/52270014) (L.Z.). M.V. is member of EMBO YIP (4053). 

We would like to acknowledge  Dr J. Cervera and Dr J.C. Pe\~nalver from IVO Foundation (Valencia, Spain).

\medskip

\noindent \textbf{Author contributions} 

\noindent V.M.P.-G. designed the research, collected and processed data, performed the fittings, developed and simulated the mathematical models. J.J.B. collected data, processed data, performed fitting tasks and developed the automatic segmentation algorithm. J.P.-B. collected and processed data. O.L.-T. collected and processed data and fitted the longitudinal volumetric data. M.V., L.X, P.G.-G., P.S.-G., E.H., R.H., performed the experiments in animal models. E.A., P.S., M.M., D.A., A.A.-R., A.F.H-M collected data. G.J. collected and processed data. M.P.-C. processed data. G.F.C., J.J.B., J.B.-B., Y.A. and A.M. developed and simulated the mathematical models. A.M.G.-V., designed the research, collected and processed data. V.M.P.-G. and G.F.C. drafted the manuscript. All authors edited and approved the final manuscript.

\medskip

%\DeclareCaptionFont{cfs}{\fontsize{8.5}{10.25}\selectfont}

\noindent \textbf{Competing interests} 

\noindent The authors declare no competing interests.

\medskip

\noindent \textbf{Ethical approval}

\noindent We have complied with all relevant ethical regulations. Human data were obtained either from public repositories (TCIA) or in the framework of several retrospective or prospective observational clinical studies approved by the corresponding institutional review boards (for details see ’Methods’). Animal care and experimental procedures were performed in accordance to the European Union and National guidelines for the use of animals in research and were reviewed and approved by the Research Ethics and Animal Wellfare Committee at Instituto de Salud Carlos III de Madrid (PROEX 244/14) (glioma cells) and in accordance with a protocol approved by the CNIO, Instituto de Salud Carlos III and Comunidad de Madrid Institutional Animal Care and Use Committee (H2030-BrM3 cells).

\medskip

\noindent \textbf{Additional information}

\noindent Supplementary information is available for this paper.

\medskip

\noindent \textbf{Correspondence and requests for materials} should be addressed to V.M. P\'erez-Garc\'{\i}a (\texttt{victor.perezgarcia@uclm.es}). 

\bigskip

\newpage

\fontfamily{cmr} \fontsize{9}{10.25} \selectfont

\noindent \textcolor{black}{\sf \bfseries \large Methods}

\medskip

\noindent \textbf{Patients and image acquisition.} Several patient datasets were included in our study. Patient subgroups 1-6 were used for the construction of the scaling laws (1,2,4,5 also for the survival studies). Patient subgroups 7-10 were used for the study of the longitudinal tumour volumetric dynamics. Overall survival (OS) was determined as the time from pretreatment imaging to death or last follow-up.
\medskip

\noindent \textbf{Breast Cancer Patients (subgroup 1).} Patients were participants of a multicenter prospective study approved by the Institutional Review Board (IRB). Written informed consent was obtained from all patients. 
The inclusion criteria were: (1) newly diagnosed locally advanced breast cancer with clinical indication of neoadjuvant chemotherapy, (2) lesion uptake higher than background, (3) absence of distant metastases confirmed by other methods previous to the request of the PET/CT for staging, and (4) breast lesion size of at least 2 cm.  54 patients (18\% lobular carcinoma, 82\% ductal carcinomas, 100\% women, age rank 25-80, median 50 years)  were included in this dataset. The TNM data were: 54\% T2, 18\% T3, 28\% T4; 28\% N0, 55\% N1, 6\% N2, 11\% N3; 100\% M0.

PET/CT examinations were performed on the same dedicated whole-body PET/CT scanner (Discovery DSTE-16s, GE Medical Systems) in three-dimensional (3D) mode. The acquisition began 60 minutes after intravenous administration of approximately 370 MBq (10 mCi) of $^{18}$F-FDG. The image voxel size was 5.47 mm $\times$  5.47 mm 
$\times$  3.27 mm, with a slice thickness of 3.27 mm and no gap between slices. 

\medskip

\noindent \textbf{Head \& Neck cancer patients (subgroup 2).} These patients were obtained from The Cancer Imaging Archive (TCIA)\cite{TCIA} Head-Neck-PET-CT collection (H\&N1 data set)  \cite{Vallieres}. This cohort was composed of patients with primary squamous cell carcinoma of the head-and-neck (stages I-IV). 76 consecutive patients from this subset satisfying the inclusion criteria (availability of pretreatment PET studies, well-defined primary tumour and lesion size larger than 2.0 cm) were included in our study. Data for the cohort were: age rank 18-84, median 62 years; 63 male, 13 female; 13 cancers of larynx, 3 hypopharynx,  11 nasopharynx, 49 oropharynx. Staging data are 3 stage II, 1 stage IIB, 26 stage III, 44 stage IVA, 2 stage IVB. TNM data were 11 T1, 19 T2, 34 T3, 12 T4; 11 N0, 16 N1, 7 N2a, 27 N2b, 13 N2c, 2 N3; 72 M0, 4 Mx. 

Eligible patients had FDG-PET scans obtained on a hybrid PET/CT scanner (Discovery ST, GE Healthcare) within 37 days before treatment (median: 14 days). A median of 584 MBq (range: 368-715) was injected intravenously. Imaging acquisition of the head and neck was performed using multiple bed positions with a median of 300 s (range: 180-420) per bed position. The slice thickness resolution was 3.27 mm for all patients and the median in-plane resolution was 3.52 $\times$ 3.52 mm$^2$ (range: 3.52-4.69). 

\medskip

\noindent \textbf{Rectal cancer (subgroup 3).} A retrospective observational study (SCALAWS: Scaling laws, shape factors and fractal measures in human cancers) was designed and approved by the IRB of the participating institutions. Inclusion criteria were: histological confirmation of advanced rectal cancer diagnosis, availability of pretreatment PET/CT and lesion size larger than 2 cm. A total of 23 rectal cancer patients (16 male, 7 female, age rank 54-80, median age 72 years) from the period October 2007 to October 2009 were included in the study. PET protocol and machine were as in subgroup 1.

\medskip

\noindent \textbf{Lung cancer (subgroup 4).} 175 patients (153 men, 22 women, age rank 41-84, median 65 years) were included in the SCALAWS study from a dataset of lung cancer patients that received surgery in the period June 2007 to December 2016. Histologies were 63 squamous cell carcinoma and 112 adenocarcinoma. Staging information was: 69 stage I, 70 stage II, 33 stage III, 3 stage IV. The N staging was 107 patients N0, 46 N1 and 22 N2. All patients had M0. PET protocol and machine were as in subgroup 1. We set the inclusion criterion that minimal lesion size should be larger than 2.0 cm.

\medskip

\noindent \textbf{Gliomas (subgroup 5).} A prospective multicenter and non-randomized study was designed (FuMeGA: Functional and Metabolic Glioma Analysis), and approved by the IRB of the participanting institutions. Informed consent was obtained from all patients. Patients were included consecutively. A basal $^{18}$F-fluorocholine PET/CT was performed in patients suspicious of glioma after magnetic resonance imaging (MRI) with an operable brain lesion and a good performance status (ECOG$\leq$2). Patients with pathologically confirmed brain glioma, and unifocal lesions of size larger than 2.0 cm were included. Our study included 44 patients from the period 2017-2019 (15 women, 29 men), age rank 23-79, median 60 years. Histologies were 32 glioblastoma IDH1wt, three glioblastoma IDH1mut, two oligodendroglioma, four difuse astrocytoma and three anaplastic astrocytoma.

PET machine was as in subgroup 1. PET adquisition was initiated 40 min after the intravenous administration of 185 MBq of $^{18}$F-Fluorocholine. A brain scan was performed starting with a low-dose CT transmission study (modulated 120 kV and 80 mA) without intravenous contrast followed by 3D emission study with an acquisition time of 20 min (one single bed), voxel size of 2.3 $\times$ 2.3 $\times$ 3.3 mm in a matrix of 128 $\times$ 128 and reconstructed using the CT images for attenuation correction and applying an iterative reconstruction algorithm. 

\medskip

\noindent \textbf{Breast cancer patients (subgroup 6).}  Pretreatment $^{18}$F-FLT PET/CT scans of patients of the ACRIN 6688  observational study available in the TCIA (ACRIN-FLT-Breast) were included in the study \cite{BFLT}. This dataset included histologically confirmed breast cancer patients (100\% women), of them 46.8\%  premenopausal and 52.2\% postmenopausal. TNM were 3\% TX, 1\% T1, 47\% T2, 34\% T3 and 14\% T4; 3\% NX, 29\% N0, 51\% N1, 11\% N2, 6\% N3; and 100\% N0.

Inclusion criteria were: (1) primary breast cancer measuring 2.0 cm or more, (2) being a candidate for neo-adjuvant chemotherapy (NAC) and surgical resection of residual primary tumour after chemotherapy, and (3) no evidence of stage IV disease. Patients received a baseline pretreatment $^{18}$F-FLT PET/CT study within 4 weeks before NAC initiation. After the injection of 2.6 MBq/kg (mean, 167 MBq; range, 110-204 MBq), a whole-body image (5-7 bed positions) was obtained at 60 min (mean, 70 min; range, 50-101 min). %All patients were scanned on calibrated and ACRIN- accredited PET/CT scanners, which included review of image quality and testing of SUVs using a uniform phantom and review of images. 
75 patients were included in the study (100\% female, age rank 22-83 y, median 50 y).

\medskip

\noindent \textbf{Brain metastasis patients (subgroup 7).} Patients were participants in the study METMATH (Metastasis and mathematics), a retrospective multicenter and non-randomized study approved by the IRB of the participanting institutions. Five patients (four women, one man; age rank 38-67, median 52 years) diagnosed of brain metastasis of a primary lung cancer with an untreated lesion with three or more consecutive MRI studies before treatment were included. Primaries were four non-small cell lung cancer and one breast luminal b cancer. A total of 16 imaging studies were included, the range of studies per patient being 3-4. 

Postcontrast T1-weighted sequence was gradient echo using 3D spoiled-gradient recalled echo or 3D fast-field echo after intravenous administration of a single-dose of gadobenate dimeglumine (0.10 mmol/kg) with a (6-8)-min delay. 

All MRI studies were performed in the axial plane with either a 1.5 T Siemens scanner, a 3 T Philips scanner and a 1 T Philips scanner. Imaging parameters were no gap, slice thickness of 1 - 1.6 mm, 0.438-0.575 mm $xy$ resolutions, and 0.8 - 1.3 mm spacing between slices.

\medskip

% The institutional review board of the participating institutions approved this retrospective analysis. All protected health information and site identifiers were removed electronically.

\noindent \textbf{Lung cancer patients (subgroup 8)}. Patients included were participants in the study SCALAMATH. Five patients (3 men, 2 women) were included. Three of them were diagnosed as adenocarcinoma and two as squamous cell carcinomas. Age rank was 60-72 years (median 68).  All of them were initially stage I tumours and progressed without treatment.  

We drew from the database of follow-up screenings in I-ELCAP between 2008 and 2019, which were performed according to a common protocol \cite{LungStaging} using low-dose CT (LDCT). Enrollment was limited to those aged 50 years or older, with a smoking history of at least 10 pack-years, no previous cancer and general good health. Participants harboring parenchymal solid or part-solid non calcified nodule with at least three or more follow-up CTs were identified according to specified criteria in the protocol.  A total of 22 imaging studies were used (range of studies per patient 3-6).

Thoracic CT scans used a 16 acquisition channels multidetector computed tomography (Siemens Emotion 16, Erlangen, Germany) with maximum section collimation of 1 mm, 0.7 mm of spacing between slices, 1 mm of slice thickness and a range of xy resolution of 0.584 - 0.783 mm. The CT scans were done at 120 kVps y 30 mAs, and less than 1 s tube rotation time. Contiguous images were reconstructed in the trans-axial plane using 1 mm thickness. Lung image sets were reconstructed with a high frequency algorithm and mediastinal image sets were reconstructed with an intermediate frequency algorithm.

Lung cancer diagnosis was made by histopathological examination of needle core biopsy or resection specimens, or by cytopathological examination of bronchoscopic or needle aspiration biopsy samples. Resected tumours were classified using the WHO classification of lung neoplasms. Adenocarcinomas were classified according to the International Association for the Study of Lung Cancer American Thoracic Society European Respiratory Society classification of lung adenocarcinoma. All lung cancer diagnoses were centrally reviewed. The tumours were staged using the International Association for the Study of Lung Cancer Staging Guidelines\cite{LungStaging}.

\medskip

\noindent \textbf{Low-grade glioma patients (subgroup 9)}.  82 patients diagnosed of grade II gliomas (biopsy/surgery confirmed astrocytoma, oligoastrocytoma or oligodendroglioma according to the WHO 2007 classification) 
and followed at the Bern University Hospital between 1990 and 2013 were initially included in the study. The study was approved by Kantonale Ethikkommission Bern (Bern, Switzerland).
% with approval number: 07.09.72.

Of that patient population, we selected patients receiving either no treatment or only surgery for which at least five post-surgery consecutive images showing tumour growth were available. Six patients initially diagnosed as grade II gliomas (age rank 29-50, mean 37 years, 4 astrocytomas and 2 oligodendrogliomas) satisfied our inclusion criteria. A total of 34 imaging studies were used, the range of studies per patient being 4 to 7 (mean 6).

\medskip

\noindent \textbf{Lung hamartoma patients (subgroup 10)}. Six patients (five men, one woman; age rank 51-63, median 58 years) diagnosed of lung hamartomas participants of the protocol SCALAMATH with longitudinal follow-up were included in the study. Imaging methods were the same as in subgroup 8. A total of 46 imaging studies were used, the range of studies per patient being 5 to 12 (mean 8).

\medskip

\noindent \textbf{PET image analysis (patient subgroups 1-6)}. At least an experienced nuclear medicine physician and an imaging engineer independently assessed the PET scans in an Advantage Windows station (v.4.). In case of discordance a third evaluator revised the images. In the visual evaluation, a PET scan was considered as positive if any uptake, higher than normal tissue background, was detected. Only positive PET scans were considered for tumour segmentation, i.e. those having a SUVmax larger than twice the background activity readings.

PET images in DICOM 
%(Digital Imaging and Communication in Medicine) 
files were imported into the scientific software package Matlab (R2018b, The MathWorks, Inc., Natick, MA, USA). The tumour PET images were manually placed in a 3D box and then semi-automatically delineated using a grey-level threshold chosen to identify the metabolic tumour volume. Segmentations were corrected manually slice by slice as in \cite{Radiology}.

 All segmentations were performed
by at least a nuclear medicine physician and an imaging engineer, both with more than 5 years of experience in tumour segmentation. In many cases one or two additional segmentations by imaging engineers were performed to verify the robustness of the methodology and to obtain consensus segmentations. 
Physiological activity contiguous with tumour uptake, as e.g. choroid plexus or skull in the brain, was manually excluded from the tumour segmentations. To avoid observer dependent biases, for those tumour histologies well separated from surrounding structures with physiological uptake an automatic segmentation algorithm was developed (see SI Section S5). 

The radiotracer standardized uptake values (SUV) were computed for each voxel using the formula
\begin{equation}
\text{SUV} = \frac{S_v \times R_S \times W}{R_{TD} \times D_F \times e^{\text{ln(2)} E_t / H_F}}.
\end{equation}
Where $S_v$ is the stored value, $R_S$ the rescaled slope, $W$ is the patient weight, $R_{TD}$ is the radiopharmaceutical injected dose and $H_F$ its half-life, $D_F$ is the decay factor, and $E_t$ is the elapsed time for each slice processed.

Global metabolic parameters were obtained, specifically the metabolic tumour volume (MTV, the volume of the VOI after segmentation) and the total lesion activity (TLA, the sum of all local SUV values over the VOI). Also relevant local metrics such as the maximum value of the SUV over the segmented lesion was stored (SUVmax). Since radiotracer uptake is very low in necrotic areas, they typically do not contribute to TLA and MTV.

\medskip

\noindent \textbf{MRI image analysis (patient subgroups 7 \& 9)}. Brain metastasis T1-weigthed images were collected in DICOM format and analysed by the same image expert (OLT, 2 years of expertise on tumour segmentation) as described for patient subgroups 1-6.  An experienced radiologist (EA) revised and validated the tumour delineation.

For subgroup 9, T2/FLAIR MRI studies were used to define the tumour volume. Radiological glioma growth was quantified by manual measurements of tumour diameters on successive MRI studies (T2/FLAIR sequences). For older imaging data available only as jpeg images we computed the volume using the ellipsoidal approximation \cite{3diam}. 

\medskip

\noindent \textbf{CT image analysis (patient subgroup 8, 10)}. Patients included were participants in the study SCALAMATH. CT images of lung cancer nodules were obtained in DICOM format. An experienced radiologist (EA) localized the lesion and then an image expert (OLT) performed the segmentations following the same methodology as with subgroups 1-7. 

\medskip

\noindent \textbf{Glioma cells.} Primary glioma cells (L0627) were kindly provided by Rosella Galli (San Raffaele Scientific Institute, Milan, Italy) and were grown in complete medium: Neurobasal (Fisher) supplemented with B27 (1:50) (Fisher); Glutamax (1:100) (Fisher); Penicillin-streptomycin (1:100) (Lonza); 0.4 \% heparin (Sigma-Aldrich); 40 ng/ml EGF and 20 ng/ml bFGF2 (Peprotech). Cells were passaged after enzymatic disaggregation using Accumax (Milipore). In order to monitor tumour growth, cells were infected with lentiviral particles expressing Fluc (pLV-Hygro-EF1A-Luciferase) (Vector-Builder) and selected in the presence of Hygromycin.

\medskip

\noindent \textbf{Mouse glioma xenografts.} Animal care and experimental procedures were performed in accordance to the European and National guidelines for the use of animals in research and were  approved by the Research Ethics and Animal Welfare Committee at Instituto de Salud Carlos III, Madrid (PROEX 244/14). Stereotactically guided intracranial injections in athymic nude Foxn1$^\text{nu}$ mice were performed by administering 1$\times$105 L0627 cells (expressing the luciferase reporter gene) resuspended in 2 $\mu l$ of culture media. The injections were made into the striatum (coordinates: A-P, $\pm$0.5 mm; M-L, +2 mm; D-V, -3 mm; related to Bregma) using a Hamilton syringe. One month after the injection we started monitoring the reporter expression in the tumours. For that, animals received and intraperitoneal injection of Luciferin (Fisher) (150mg/Kg) and the Luciferase activity was visualized in an IVIS Spectrum in vivo imaging system (Perkin Elmer). The total flux (in photons per second) was measured to assess tumour growth.

\medskip

\noindent \textbf{Animal studies with H2030-BrM3 cells.}  The human lung adenocarcinoma brain tropic model H2030-BrM \cite{Nguyen}  was injected into the heart of nude mice in order to induce the formation of brain metastasis from systemically disseminated cancer cells. Brain colonization and growth of metastases were followed using non-invasive bioluminescence imaging 
 since BrM cells express luciferase. Upon administration of the substrate D-luciferin, bioluminescence generated by cancer cells was measured over the course of the disease. The increase in photon flux values is a well established correlate of tumour growth in vivo \cite{Nguyen,Valiente}.The experiments were performed in accordance with a protocol approved by the CNIO, Instituto de Salud Carlos III and Comunidad de Madrid Institutional Animal Care and Use Committee. Athymic nu/nu (Harlan) mice of 4-6 weeks of age were used. Brain colonization assays were performed by injecting 100 $\mu l$ PBS into the left ventricle containing 100,000 cancer cells. Anesthetized mice (isofluorane) were injected retro-orbitally with d-luciferin (150 mg/kg) and imaged with an IVIS Xenogen machine (Caliper Life Sciences). Bioluminescence analysis was performed using Living Image software, version 3.

\medskip

\noindent \textbf{Cell culture.}  H2030-BrM3 (abbreviated as H2030-BrM) was cultured in RPMI1640 media supplemented with 10\% 
FBS, 2 mM l-glutamine, 100 IU ml\textsuperscript{-1} penicillin/ streptomycin and 1 mg ml\textsuperscript{-1} amphotericin B.

\medskip

\noindent \textbf{Statistical analysis.} 
Linear regressions of the $\log\left(\text{MTV}\right)$ versus $\log\left(\text{TLA}\right)$ to construct the scaling laws were performed using the Matlab Statistics and Machine Learning toolbox command \texttt{fitlm}. In Figure \ref{explosion}, the nonlinear fittings were carried out by fixing the optimum $\beta$ for all patients of the same cancer type allowing only for the personalization of the growth parameter $\alpha$. Thus, for every set of $N$ patients with the same cancer type having a total of $M (> 3N)$ data points we fitted the $N$ values of $\alpha$. For each cancer type, the value for $\beta$ that was used was the one providing the smallest mean squared error. To fit the longitudinal tumour volumetric dynamics to the model (\ref{themodel}) with different values for $\beta=3/4, 1$,  $5/4$ and optimum we used the Matlab function \texttt{fmincon}.
\par
The Harrell's concordance index (c-index) \cite{Harrell} was computed to evaluate the model's capacity to discriminate patient subgroups with different survival. We computed the c-index for each possible threshold $\alpha$ in the scaling law $\log \text{TLA} = \log(\alpha) + \frac{5}{4} \log \text{MTV}$ or the metabolic variables (TLA, MTV) splitting the patient population into two groups (values above and below the line) and searched for the non-isolated significant values ($p<0.1$) obtaining the highest value of the c-index. Kaplan-Meier curves were constructed to compare both populations and the log-rank two-tailed test used to compute the c-index. When either no curves with $p<0.1$ were found or the best c-index obtained was below the value 0.7, the variable under study was considered to be unable to classify patients accurately in terms of survival.
\medskip

\noindent \textbf{Data availability.} 
Source data for figures 1, 2 and 4 are available for this paper. All other data that support the plots within this paper and other findings of this study are available from the corresponding author upon reasonable request.

\medskip
\noindent \textbf{Code availability.}  The mesoscopic simulator code is available for download from \texttt{http://matematicas.uclm.es/molab/DiscrSimulator1.zip}

\renewcommand{\refname}{}
\fontfamily{lmss} \fontsize{8.0}{10.25} \selectfont

\bigskip
\noindent{\bf\normalsize References}
\par


\vspace*{-3mm}
\begin{thebibliography}{99}

\bibitem{Kleiber} Kleiber, M. Body size and metabolism. \emph{Hilgardia} \textbf{6,} 315-351 (1932).

\bibitem{WestBook} West, G. \emph{Scale: The Universal Laws of Growth Innovation, Sustainability, and the Pace of Life in Organisms, Cities, Economies and Companies} (Penguin Press, New York, 2017).

\bibitem{WBB} West, G. B., Brown, J. H. \& Enquist, B. J.  A general model for the origin of allometric scaling laws in biology. \emph{Science} \textbf{276,} 122-126 (1997).

%\bibitem{Kolo} Kolokotrones., T, Deeds, E.J., \& Fontana, W. Curvature in metabolic scaling. \emph{Nature} \textbf{464,} 753-756 (2010).

%\bibitem{WestBrown} West GB, Brown JH. Life's universal scaling laws. \emph{Phys. Today}  \textbf{57,} 36-42 (2004).

\bibitem{R1} Savage, V. M. et al.
%Gillooly JF, Woodruff WH, West GB, Allen AP, Enquist BJ, Brown JH. 
The predominance of quarter-power scaling in biology. \emph{Funct. Ecol.} \textbf{18,} 257-282 (2004).

\bibitem{Reich} Reich, P. B., Tjoelker, M. G., Machado, J. L. \& Oleksyn, J. Universal scaling of respiratory metabolism, size and nitrogen in plants. \emph{Nature} \textbf{439,} 457-461 (2006).

\bibitem{R4} Enquist, B. J. et al.
%Allen AP, Brown JH, Gillooly JF, Kerkhoff AJ, Niklas KJ, Price CA, West GB. 
Biological scaling: does the exception prove the rule? \emph{Nature}. \textbf{445,} E9-10 (2007).

\bibitem{UniversalLaw} Guiot, C., Degiorgis, P. G., Delsanto, P. P., Gabriele, P. \& Deisboeck, T. S. Does tumour growth follow a universal law? {\em J. Theor. Biol.} \textbf{ 225,} 147-151 (2003).

\bibitem{Herman} Herman, A. B., Savage, V. M. \& West, G. B. A quantitative theory of solid tumor growth, metabolic rate and vascularization. {\em PLoS One} \textbf{ 6,} e22973 (2011).

\bibitem{MetabScal} Milotti, E., Vyshemirsky, V., Sega, M., Stella, S. \& Chignola, R. Metabolic scaling in solid tumours. {\em Sci. Rep.} \textbf{3,} 1938 (2013).

\bibitem{Palm} Palm, W. \& Thompson, C. B. Nutrient acquisition strategies in mammalian cells. {\em Nature} \textbf{546,} 234-242 (2017).

\bibitem{Zhu2019} Zhu, J. \& Thompson, C. B. Metabolic regulation of cell growth and proliferation. {\em Nat. Rev. Mol. Cell Biol.} \textbf{20,} 436-450 (2019).

%\bibitem{Thompson} Vander Heiden, M. G., Cantley, L. C. \& Thompson, C. B. Understanding the Warburg effect: the metabolic requirements of cell proliferation. {\em Science} \textbf{ 324}, 1029-1033 (2009).

\bibitem{Choline} De Grado, T. R., Reiman, R. E., Price, D. T., Wang, S. \& Coleman, R. E. Pharmacokinetics and radiation dosimetry of $^{18}$F-fluorocholine. \emph{J. Nucl. Med.}, \textbf{43,} 92-96 (2002).

\bibitem{FLT} Barwick, T., Bencherif, B., Mountz, J. M. \& Avril, N. Molecular PET and PET/CT imaging of tumour cell proliferation using F-18 fluoro-L-thymidine: a comprehensive evaluation. \emph{Nucl. Med. Commun.}, \textbf{30,} 908-917 (2009).

%\bibitem{Bettencourt2013} Bettencourt, L. M. A. The origins of scaling in cities. {\em Science} \textbf{340,} 1438-1441 (2013).

%\bibitem{Li2017} Li, R. et al. Simple spatial scaling rules behind complex cities. {\em Nat. Commun.} \textbf{8,} 1841 (2017).

\bibitem{DeLong} DeLong, J. P., Okie, J. G., Moses, M. E., Sibly, R. M. \& Brown, J. H. Shifts in metabolic scaling, production, and efficiency across major evolutionary transitions of life. {\em Proc. Natl Acad. Sci. USA} \textbf{107,} 12941-12945 (2010).

\bibitem{DePalma} De Palma, M., Biziato, D. \& Petrova, T. Microenvironmental regulation of tumour angiogenesis. {\em Nat. Rev. Cancer} \textbf{17,} 457-474 (2017).

%\bibitem{Atavism1} Bussey, K.J., Cisneros, L.H., Lineweaver, C.H., \& Davies, P.C.W. Ancestral gene regulatory networks drive cancer.
%\emph{Proc. Natl Acad. Sci. USA}  \textbf{114,} 6160-6162 (2017).

%\bibitem{Atavism2} Trigos, A.S., Pearson, R.B., Papenfuss, A.T., Goode, D.L. Altered interactions between unicellular and multicellular genes drive hallmarks of transformation in a diverse range of solid tumors. \emph{Proc. Natl Acad. Sci. USA} \textbf{114,} 6406-6411 (2017).

\bibitem{WestLaw} West, G. B., Brown, J. H. \& Enquist, B. J. A general model for ontogenetic growth. \emph{Nature} \textbf{413,} 628-631 (2001).

%\bibitem{Martin} Martin, M. Researchers suggest that universal law governs tumour growth. \emph{J. Natl. Cancer Institute} \textbf{95,} 704-705 (2003).

\bibitem{Gerlee} Gerlee, P. The model muddle: in search of tumor growth laws. \emph{Cancer Res.} \textbf{73,} 2407-2411 (2013).

\bibitem{RB2013} Rodriguez-Brenes, I. A., Komarova, N. L. \& Wodarz, D. Tumor growth dynamics: insights into evolutionary processes. \emph{Trends Ecol. Evol.} \textbf{28,} 597-604 (2013).

\bibitem{Benzecry}  Benzekry, S. et al.
%Lamont, C., Beheshti, A., Tracz, A., Ebos, J. M. L., Hlatky, L. \& Hahnfeldt, P. 
Classical mathematical models for description and prediction of experimental tumour growth. \emph{PLoS Comput. Biol.} \textbf{10,} e1003800 (2014).

\bibitem{Talkington} Talkington, A. \& Durret, R. Estimating tumour growth rates in vivo. \emph{Bull. Math. Biol.} \textbf{77,} 1934-1954 (2015).

\bibitem{Gengenbacher} Gengenbacher, N., Singhal, M. \& Augustin, H. G. Preclinical mouse solid tumour models: status quo, challenges and perspectives. {\em Nat. Rev. Cancer} \textbf{17,} 751-765 (2017).

%\bibitem{Nowak-Sliwinska} Nowak-Sliwinska, P. et al. Consensus guidelines for the use and interpretation of angiogenesis assays. \emph{Angiogenesis} \textbf{21,} 425-532 (2018).

%\bibitem{Suecos} Stensjoen, A. L., Solheim, O., Kvistad, K. A., Haberg, A. K., Salvesen, O. \& Berntsen, E. M. Growth dynamics of untreated glioblastomas in vivo. \emph{Neuro-Oncology}, \textbf{17,} 1402-1411 (2015).

\bibitem{Mandonnet} Mandonnet, E. et al. Continuous growth of mean tumour diameter in a subset of grade II gliomas. \emph{Ann. Neurol.} \textbf{53,} 524-528 (2003).

\bibitem{VanHavenbergh} Van Havenbergh, T., Carvalho, G., Tatagiba, M., Plets, C. \& Samii, M. Natural history of petroclival meningiomas. \emph{Neurosurgery} \textbf{52,} 55-64 (2003).

\bibitem{Heesterman} Heesterman, B. L. et al. Mathematical models for tumour growth and the reduction of overtreatment. \emph{J. Neurol. Surg. B} \textbf{80,} 72-78 (2019).

%\bibitem{LungStaging}  Henschke, C.I. International Early Lung Cancer Action Program: Enrollment and Screening Protocol. Accessed at http://www.ielcap.org/sites/default/files/I-ELCAP-protocol.pdf on 14 March 2019.

\bibitem{Nguyen} Nguyen, D.X. et al. WNT/TCF signaling through LEF1 and HOXB9 mediates lung adenocarcinoma metastasis. {\em Cell} \textbf{138,} 51-62 (2009).

\bibitem{STM} Gargini, R. et al. The IDH-TAU-EGFR triad defines the neovascular landscape of diffuse gliomas. {\em Sci. Trasl. Med.} (to appear) (2020).

\bibitem{Hallatschek} Hallastschek, O. \& Fisher, D. S. Acceleration of evolutionary spread by long-range dispersal. {\em Proc. Natl Acad. Sci. USA} \textbf{111,} E4911-E4919 (2014).

\bibitem{Deforet} Deforet, M., Carmona-Fontaine, C., Korolev, K. S. \& Xavier, J. B. Evolution at the edge of expanding populations. {\em Am. Nat.} \textbf{194,} 291-305 (2019).
 
\bibitem{ED1} Waclaw, B., Bozic, I., Pittman, M. E., Hruban, R. H., Vogelstein, B. \& Nowak, M. A. A spatial model predicts that dispersal and cell turnover limit intratumour heterogeneity.  Nature \textbf{525,} 261-264 (2015).

\bibitem{ED2}  Komarova, N. L. Spatial interactions and cooperation can change the speed of evolution of complex phenotypes. \emph{Proc. Natl Acad. Sci. USA} \textbf{111,} 10789-10795 (2014).

\bibitem{Michor} Durrett, R., Foo, J., Leder, K., Mayberry, J. \& Michor, F. Evolutionary dynamics of tumor progression with random fitness values. {\em Theor. Popul. Biol.} \textbf{78,} 54-66 (2010).

\bibitem{McFarland} McFarland, C. D., Mirny, L. A., Korolev, K. S. Tug-of-war between driver and passenger mutations in cancer and other adaptive processes. {\em Proc. Natl Acad. Sci. USA} \textbf{111,} 15138-15143 (2014).

\bibitem{Anderson} Robertson-Tessi, M., Gillies, R. J., Gatenby, R. A., Anderson, A. R. Impact of metabolic heterogeneity on tumor growth, invasion, and treatment outcomes.
 \emph{Cancer Res.} \textbf{75,} 1567-1579 (2015).

%\bibitem{Radiology} P\'erez-Beteta, J. et al. \emph{Tumor Surface Regularity at MR Imaging Predicts Survival and Response to Surgery in Patients with Glioblastoma}. Radiology \textbf{288}, 218-225 (2018).

%, Molina-Garc\'ia, D., J.A. Ort\'iz-Alhambra, A. Fern\'andez-Romero, B. Luque, E. Arregui, M. Calvo, J.M. Borr\'as, B. Mel\'endez, A. Rodr\'iguez de Lope, R. Moreno de la Presa, L. Iglesias Bayo, J.A. Barcia, J. Martino, C. Vel\'asquez, B. Asenjo, M. Benavides, I. Herruzo, A. Revert, E. Arana, V.M. P\'erez-Garc\'ia. 
\setcounter{mine}{\value{enumiv}}

\end{thebibliography}

\vspace*{-5mm}
\begin{thebibliography}{99}

\setcounter{enumiv}{\value{mine}}

%\bibitem{Previous} Garc\'{\i}a Vicente, A. M. et al.
%Cruz Mora MÃÂÃÂ, LeÃÂÃÂ³n MartÃÂÃÂ­n AA, MuÃÂÃÂ±oz SÃÂÃÂ¡nchez M del M, Relea Calatayud F, Van GÃÂÃÂ³mez LÃÂÃÂ³pez et al. 
%Glycolytic activity with 18F-FDG PET/CT predicts final neoadjuvant chemotherapy response in breast cancer. \emph{Tumor Biol.} {\bf 35,} 11613-11620 (2014).

\bibitem{TCIA} Clark, K. et al. 
%Vendt B, Smith K, Freymann J, Kirby J, Koppel P, Moore S, Phillips S, Maffitt D, Pringle M, Tarbox L, Prior F.
 The cancer imaging archive (TCIA): maintaining and operating a public information repository. \emph{J of Dig Imag}  \textbf{26,} 1045-1057 (2013).

\bibitem{Vallieres} Vallieres, M. et al. Radiomics strategies for risk assessment of tumour failure in head-and-neck cancer. \emph{Sci. Rep.} \textbf{7,} 10117 (2017).

\bibitem{BFLT} Kostakoglu, L. et al.
%Duan F, Idowu MO, Jolles PR, Bear HD, Muzi M, Cormack J, Muzi JP, Pryma DA, Specht JM, Hovanessian-Larsen L, Miliziano J, Mallett S, Shields AF, Mankoff DA; 
A phase II study of 3'-Deoxy-3'-18F-Fluorothymidine PET in the assessment of early response of breast cancer to neoadjuvant chemotherapy: results from ACRIN 6688. \emph{J. Nucl. Med.}  \textbf{56,} 1681-1689 (2015). 

\bibitem{LungStaging} Detterbeck, F. C., Boffa, D. J., Kim, A. W., Tanoue, L. T. The eighth edition lung cancer stage classification. \emph{Chest} 
\textbf{ 151,} 193-203 (2017).

\bibitem{PVE2007} Soret, M., Bacharach, S. L., Buvat, I. Partial-volume effect in PET tumor imaging. \emph{J. Nucl. Med.} \textbf{48,} 932-945 (2007).


\bibitem{3diam} Pallud, J. et al. Prognostic value of initial magnetic resonance imaging growth rates for world health organization grade II gliomas.  \emph{Ann. Neurol.} \textbf{60,} 380-383 (2006).

%\bibitem{Nguyen} Nguyen, D.X. et al. WNT/TCF signaling through LEF1 and HOXB9 mediates lung adenocarcinoma metastasis. {\em Cell} \textbf{138,} 51-62 (2009).  

\bibitem{Valiente} Valiente M. et al. Serpins promote cancer cell survival and vascular co-option in brain metastasis. {\em Cell}  \textbf{156,} 1002-1016 (2014).

\bibitem{Radiology} P\'erez-Beteta J, et al. 
tumour Surface Regularity at MR Imaging Predicts Survival and Response to Surgery in Patients with Glioblastoma.
\emph{Radiology} 288(1):218-225 (2018).


\bibitem{Harrell} Harrell, F. E. et al. Evaluating the yield of medical tests. \emph{JAMA-J. Am. Med. Assoc.} \textbf{247,} 2543-2546 (1982).

%\bibitem{FKmodel} Murray, J. \emph{Mathematical Biology.} Third Edition. (Springer, 2002). 

\end{thebibliography}
\end{document}